\documentclass[aps,prl,twocolumn,superscriptaddress,groupedaddress,showpacs]{revtex4-1}
\usepackage{amsmath}
\usepackage{graphicx}% Include figure files
\usepackage{dcolumn}% Align table columns on decimal point
\usepackage{bm}% bold math
\usepackage[mathlines]{lineno}% Enable numbering of text and display math
\usepackage{tensor}
\usepackage{mathtools}
\usepackage{natbib}
\usepackage{enumerate}
\usepackage{slashed}
\usepackage{comment}
\usepackage{amsfonts,amssymb,amsthm,array,amsmath,mathrsfs}
\usepackage{relsize}
\usepackage{cancel}
\usepackage{mathtools}
\usepackage[T1]{fontenc}
\usepackage{fancyhdr}
\usepackage{soul}

\usepackage[english]{babel}
\usepackage[utf8]{inputenc}

\usepackage{marginnote}
\usepackage[normalem]{ulem}

\usepackage{subfigure}

\newcommand{\nn}{\nonumber}
\def\tr{\mbox{tr}}

\newcommand{\eq}[1]{(\ref{#1})}

\newcommand{\la}{\label}
\newcommand{\ba}{\begin{align}}
\newcommand{\ee}{\end{equation}}
\newcommand{\be}{\begin{equation}}

\def\12{\frac{1}{2}}
\newcommand{\p}{\partial}
\newcommand{\en}{\end{align}}
\newcommand{\e}{\varepsilon}

\usepackage{color}

\begin{document}

\title{Chern-Simons modification of 
 Fluid Mechanics}
 \author{P.B.~Wiegmann}
 \affiliation{
 Kadanoff Center for Theoretical Physics, University of Chicago,
 5640 South Ellis Ave, Chicago, IL 60637, USA
}

%\author{Charlie Author}
% \homepage{http://www.Second.institution.edu/~Charlie.Author}
%\affiliation{
% Second institution and/or address\\
% This line break forced% with \\
%}%
%\affiliation{
% Third institution, the second for Charlie Author
%}%
%\author{Delta Author}
%\affiliation{%
% Authors' institution and/or address\\
% This line break forced with \textbackslash\textbackslash
%}%

\date{\today}% It is always \today, today,
             %  but any date may be explicitly specified

%.%%%%%%%%%%%%%%%%%%%%%%%%%%%%%%%%%%%%%%%%%%%%
%\documentclass[11pt,a4paper]{article}
%    \usepackage{jheppub}
     %\documentclass[aps,prl,twocolumn,superscriptaddress,groupedaddress,showpacs]{revtex4-1}
 % for review and %submission
%\documentclass[aps,preprint,superscriptaddress,groupedaddress,nofootinbib]{revtex4-1}
 % for double-spaced preprint
%\documentclass[aps,prl,twocolumn,superscriptaddress,groupedaddress,nofootinbib]{revtex4-1}%\documentclass[10pt,twocolumn,aps,prl]{revtex4-1}
%%%%%%%%%%%%%%%%
%\usepackage[style=phys,url,doi,eprint]{biblatex}\usepackage{amsmath,amssymb}
%\usepackage{amsfonts}

%%%%%%%%%%%%%%%%%%%%%%

%\begin{document}

%\setcounter{secnumdepth}{-1} 

\begin{abstract}

We show that the hydrodynamics of a perfect fluid admits a natural modification that incorporates a chiral gravitational anomaly (also known as a mixed gauge-gravity anomaly) alongside the  chiral current anomaly. This modification introduces in parallel a gravitational and gauge Chern-Simons terms in a manner analogous to the Jackiw-Pi Chern-Simons modification  of gravity \cite{JPi}, and  features an axion  in fluid mechanics. As a result, flows of a spinless perfect fluid acquire spin through motion, with the spin being equal to the  fluid helicity. Furthermore, spacetime curvature induces an analogue of the Mathisson-Papapetrou force and a geometric counterpart of the chiral magnetic effect.

\end{abstract}
%\notoc
%\begin{document} 
 \maketitle
%\flushbottom

 \paragraph{1. Introduction.}
In 2003 paper  \cite{JPi}, Jackiw and Pi  suggested a  modification  of the  Einstein-Hilbert
action of general relativity by the gravitational analog of the  Chern-Simons form \begin{align}
K_G^\mu=-\tfrac 1{(2\pi)^2}\e^{\mu\nu\lambda\sigma}\tr\left( \tensor{\Gamma}{_\nu}\p_\lambda
\tensor{\Gamma}{_\sigma}+\tfrac 23 \tensor{\Gamma}{_\nu}
\tensor{\Gamma}{_\lambda} \tensor{\Gamma}{_\sigma}\right)\,\la{109}
\end{align}  
coupled with the axion field
\(\Theta\). The added term reads
\begin{align}
-\tfrac k{24}\int  K_G^\mu\,\p_\mu\Theta\,.\la{2111}
\end{align} Here \(\tensor{\Gamma_\mu:=(\Gamma}{^\alpha_\beta})_\mu\)
is the affine
connection treated as  \(O(4) \)-matrix and \(k\) is a pseudo-scalar parameter.  Jackiw and Pi  called it the "Chern-Simons modification
of general relativity". The deformation brakes the \(CP\)-symmetry.  

An equivalent form of the added term \eq{2111} is the first spacetime
Pontryagin  form  \(p_1(M)=-\tfrac 12\tfrac 1{(2\pi)^2}\tr (R\wedge R)\)
 (coupled with \(\Theta\))
\be 
 \tfrac {k}2\tfrac {1}2 \tfrac 1{(2\pi)^2}\tfrac
1 {24}\int\Theta\,\, \,\tr\left(
 R\wedge R\right)\,,\la{1111}\ee
 where   \(R=\tfrac 12 (\tensor{R}{^\alpha_\beta})_{\mu\nu} dx^\mu\wedge dx^\nu
\)  is the matrix  valued Riemann curvature 2-form and \((\tensor{R}{^\alpha_\beta})_{\mu\nu} \) is the Riemann tensor treated as \(O(4)\) matrix. 

In this paper, we argue that a similar modification is relevant to fluid mechanics \cite{FootnoteNair}. In this case, the Chern-Simons form is defined on the cotangent bundle of the fluid's phase space. We  argue that a modified fluid action functional  agrees with the kinematics of the  liquid  state comprised by \(k\)-species of chiral (Weyl's) fermions, giving a realization of the chiral gauge and  gravitational anomaly in fluid mechanics \cite{Ref}. 
The paper extends the analysis of our  recent paper \cite{W}, where we considered a modification  of the
 hydrodynamics of the perfect fluid  by the multivalued Wess-Zumino-Novikov term.
Below, we obtain both modifications in a unified framework with the emphasis to the gravitational part and interpretation of the 
 the axion field  \(\Theta \), which we refer to as a {\it chiral phase}. In the fluid context it gains a clear physical interpretation also discussed in  \cite{Abanov_Cappelli, W}.

Before we proceed, two remarks are in order.  

Our treatment is classical, but it relies on the compactness of the fluid phase space. Compactness is a feature of semiclassical fluids, manifested  in Onsager's quantization of circulation: In semiclassical fluids, circulation is quantized in units of the Planck constant  \(2\pi\hbar\) \cite{Onsager}.
 In the classical fluid, it suffices to assume that  circulation is bounded away from zero. This introduces a new scale of the  dimension  of the Planck constant [Momentum] x [Length],  absent in classical fluid dynamics. For uniform notation, we set a minimal non-zero circulation to \(2\pi\), which corresponds to   \(\hbar=1\) in a semiclassical fluid. In these units, the deformation parameter  \(k\) is an integer and the chiral phase \(\Theta\) takes values on a circle  \([0,2\pi)\).
 
 The second remark concerns the class of flows. The Chern-Simons modification is only possible if the 4-fluid vorticity is a non-degenerate 2-form. This means that all four spacetime components of the fluid momentum (three components of the space-like momentum and the energy density) are  independent. Flows of this type are often called non-homentropic. These are the flows we consider. Such flows are also baroclinic, where the gradient of pressure and the gradient of fluid density are non-collinear. A contraposed class of flows where the Chern-Simons modification is not applicable are homentropic flows. These flows are barotropic. The 4-vorticity of these flows  is degenerate. We discuss this aspect in par.  3.
 
We employ the covariant formulation of hydrodynamics of Lichnerowicz and Carter \cite{Lichnerowicz, Carter}. This approach is technically simpler and less specialized in the relativistic setting, although our final formulas also apply to the Galilean fluid.

We remark that the equations of fluid dynamics must maintain covariance under the action of the gauge group and the group of spacetime diffeomorphisms and cannot be deformed at will. Specifically,  the Hamilton principle we use here can serve as a concise criterion for ensuring this property. Carter's  canonical approach to fluid mechanics is powerful and elegant, but  is not broadly familiar. We briefly review the main points in par. 2.

We begin by discussing the chiral phase \(\Theta\) and introduce the notion of fluid momentum (pars. 2 and 3), using the perfect fluid as an example. In pars. 4 and 5, we introduce the modified fluid action. In the remaining part of the paper, we study the fluid's equations of motion  and some physical consequences introduced by the modification, such as the {\it chiral anomaly} (par. 6) and spin current (par. 8).  \\

\paragraph{2. Chiral phase and fluid momentum.}  In this paragraph we briefly review the covariant canonical approach to fluid dynamics developed by Lichnerowicz and Carter. We refer the reader to two recent reviews  \cite{Gourgoulhon,Markakis}.  Using the perfect fluid as an example, we introduce canonical Eulerian fields:  the particle number current \(n^\mu\) and the fluid canonical momentum \(\pi_\mu\), conjugated via  the fluid action. We also introduce  the fluid Hamilton functional, the Legendre transform of the fluid action.  This gives us a hydrodynamic setting and the framework for admissible extension of hydrodynamics. 

We characterize the fluid motion
by   particle  4-current
\(n^\mu= (n^0, n^i)\), whose  time-like component
is the particle number density, and  the space-like component is the
particle flux through the space-like surface element  normal to  the direction \( i \) \cite{FootnoteN}. Given the particle 4-current \(n^\mu\)
and the spacetime metric we define the particle number density  \(n\) at the rest frame  as a Lorentz scalar given by  \be
n^\mu n_\mu=-n^2\,,\la{6}\ee 
and
the 4-velocity\be u^\mu=n^\mu/n,\quad u^\mu u_\mu=-1\,,\la{111}
\ee
 a unit tangent vector to the flow, defining a flow vector field. 

The gauge symmetry will be better revealed if we assume that the fluid particles  are identically electrically charged and  introduce an external electromagnetic vector potential \( A_\mu\). Such fluid is not neutral.  We assume that it is neutralized  by an oppositely charged reservoir, a medium of  particles having no inertia. We denote the particle 4-current of the reservoir by \(\bar n^\mu\) and the reservoir particle  density  by \(\bar n\).    The fluid  energy depends on the reservoir density \(\bar n\), but  does not depend on its motion, hence on its  current \(\bar  n^\mu\).  Such particles are referred to as {\it spectator medium}.

The vector  \(n^\mu\)   and the {\it advection scalar}, the ratio \(\bar n/n\), are the arguments of the  fluid Lagrangian.  
In the relativistic case, the Lagrangian density  is obtained from the  energy density \(e(n, \bar n/n)\) at rest by treating the first argument  \(n\) as \(\sqrt{-n^\mu
n_\mu}\) 
 \be L_0[n^\mu]=-e\left(\sqrt{-n^\mu
n_\mu},\bar n/n\right)\,.\la{66}
\ee 
A coupling with the external gauge potential  extends the Lagrangian density as
\be L_0=-e+ A_\mu^\Theta n^\mu-A_\mu\bar
n^\mu,\quad A_\mu^\Theta =A_\mu+\p_\mu\Theta \,.\la{101}\ee
It is important to note that the vector potentials \(A^\Theta\) and \(A\)  coupled to the fluid and the reservoir appear in different gauges. They define yet another advection parameter, the chiral phase \(\Theta\) as a conjugate to the divergence of the  particle number current \(\p_\mu n^\mu\).  While the chiral phase and  the current of spectator medium  eventually drop out of the Euler equations and are often disregarded from the start in the analysis of the perfect fluid, they can not be ignored  when we deform the action. Analogous to the theta-vacuum in gauge field theories, \(\Theta\) as well as \(\bar n^\mu\) take on physical significance as  true dynamical fields. 
 Given the Lagrangian density, we define the action \({\cal S}_0=\int L_0\), and  the canonical fluid momentum as a  conjugate to the particle number current  \cite{Carter}
\begin{align}
\pi_\mu:=\delta{\cal S}_0/\delta n^\mu\,.
\end{align}
It reads
\be \pi_\mu=p_\mu+A_\mu^\Theta,
\ee
where \be p_\mu=(\p e/\p n) u_\mu\la{8}\ee
is the  kinematic fluid momentum. Particles of the reservoir do not have kinematic momentum since the Lagrangian \eq{66} does not depend on their velocity \cite{FootnoteE}.

The relation
\be p_\mu n^\mu=-n\p e/\p n
\la{w}\ee
followed from \eq{111} and \eq{8} relates  the kinematic momentum and the particle number current via a given function of \(n\) and \(\bar n\). This relation is the covariant form of the equation of state.

Canonical fluid momentum, a covector \(\pi_\mu\), forms the {\it momentum phase space} (a cotangent  bundle).  Correspondingly,  the Legendre transform of the action with respect to \( n^\mu \), a covariant version of the Hamiltonian, is  referred to as a Hamilton functional, is a functional in the  momentum phase space and a function of \( \bar n/n \). 
\be\Lambda_0[\pi]:=\int  [\pi_\mu n^\mu -A_\mu\bar n^\mu]-{\cal S}_0\,\ee 
 A simple calculation yields the classical result   of  Ref.  \cite{Whitham,Schutz}
   \be\Lambda_0[\pi]=-\int P \,,\la{11}\ee
   where  \(-P=e-n\p e/\p n\) is the (negative) fluid pressure, and where  the integration goes over the spacetime manifold \(M^4\).
   
We remark that in the Hamilton functional of the perfect
fluid   \(\pi\)   and \(\Theta\) appear in the gauge invariant combination
of the  form of the  kinematic momentum \(p_\mu\).  
This property, however, no longer holds under  the modification we consider below. 
We summarize the momentum phase space specification of fluid mechanics: given the fluid pressure as
a function of \(p_\mu\) (and \(\bar n/n\)) we determine the particle number current as
\begin{align}
n^\mu:=-\p P/\p p_\mu \la{171}
\end{align}
and use the relation between \(n^\mu\) and \(p_\mu\)  to compliment the  equations of motion
 written in terms of  canonical variables \(n^\mu\) and \(\pi_\mu\).

Once the Hamiltonian functional is set, the equations of motion follow from fundamental fluid symmetry groups: the group of diffeomorphisms of spacetime and the group of gauge transformations. Then the equations of motion appear covariant under the actions of these symmetry groups. This is a rigid requirement, making some modifications attempted in the literature invalid. However, a modification of the Hamiltonian functional \eq{11} within the already set canonical structure discussed below ensures the required covariance (cf., Ref. \cite{Nair}, where a similar a approach was utilized).
\\
\paragraph{3. Eulerian equations of motion.}
One way to obtain  fluid equations of motion is to  cast them  in the form of Noether conservation laws of energy-momentum and  particle number,  generated by  diffeomorphisms of spacetime and by the gauge transformations. 

First, the gauge transformation \(A_\mu\to A_\mu+\p_\mu\alpha\) yields the neutrality condition \(\p_\mu(n^\mu-\bar n^\mu)=0\).
Then we turn to  spacetime
diffeomorphisms \(x^\mu\to x^\mu+\epsilon^\mu(x)\). This transformation should be performed over the metric, volume, and the external gauge field  at  fixed Eulerian fields (including the  chiral phase).  The transformation of the   Hamilton functional 
   \(\Lambda_0\) in  \eq{11},
is well known. 
%(we refer to \cite{Gourgoulhon,Markakis} and references therein).
  With the already obtained neutrality condition taken into account, the result reads\begin{align}
\delta_\epsilon\Lambda_0=\int \epsilon^\nu [\p_\mu \tensor{T}{^\mu_\nu}+n^\mu F_{\mu\nu}+\p_\nu
\Theta\,(\p _\mu n^\mu)]\,.\la{211}
\end{align}
where \begin{align}
\tensor{T}{^\mu_\nu}=n^\mu p_\nu+P\,\tensor{\delta}{^\mu_\nu}\,\la{16}
\end{align}
is the  momentum-stress-energy tensor of the perfect fluid and \(F_{\mu\nu}=\p_\mu A_\nu-\p_\nu A_\mu\) is the field tensor.

The remaining part  of the gauge   transformation \(\Theta\to\Theta+\delta\Theta\)  (performed at a fixed \(\pi_\mu\)) generates the divergence of the Noether current  \(\p_\mu n^\mu\) and the gauge symmetry  yields it nullification via  the continuity equation
\be \p_\mu n^\mu=0\,. \label{c}\ee 
It nullifies the last term in \eq{211}, thereby eliminating
\(\Theta\) along with it.
 We obtain  the  familiar equation equating the force the fluid exerts upon itself with the Lorentz force
 \begin{align}
&\p_\mu \tensor{T}{^\mu_\nu}=-n^\mu F_{\mu\nu} \,.\la{241}
\end{align}
 Five equations \eq{c} and \eq{241} supplemented by the equation of state in the form  \eq{w} or \eq{171} is a complete set of equations of motion. One may complement them by the advection  of the chiral phase and  the  spectator density per particle \(\bar n/n\).
\\
\paragraph{4. Fluid phase space.} Vorticity 2-form  is defined as \be\Omega=\tfrac
 12\Omega_{\mu\nu} dx^\mu\wedge dx^\nu\,,\la{1409}\ee
  where \(\Omega_{\mu\nu}=\p_\mu
\pi_\nu-\p_\nu \pi_\mu\) is the 4-vorticity tensor. If all four components
of the 4-momentum
\(\pi_\mu\),  are independent, the vorticity form is non-degenerate: \({\rm det}\,\Omega_{\mu\nu}\) is nowhere zero. Then, in virtue of the identity \(\sqrt{{\rm det}\,\Omega_{\mu\nu}}=\12\epsilon^{\mu\nu\alpha\beta}\Omega_{\mu\nu}\Omega_{\alpha\beta}\), the 4-form \be p_1(\Omega)=-\tfrac 12\tfrac{1}{(2\pi)^2}\Omega\wedge\Omega\label{20}\ee
%is nowhere zero. It defines the first Pontryagin form associated with the momentum phase space. The form is  realized on the spacetime from which one has to remove    the world surfaces of vortex tubes. On such  manifold one defines the first Pontryagin number 
%\be
% p_1(\Omega)=-\tfrac 12\tfrac{1}{(2\pi)^2}\int \Omega\wedge\Omega\la{C}\ee taking values in even integers. This spacetime topological invariant  is a necessary element of our construct. We discuss its geometric meaning in the end of par. 8.

Flows with a non-degenerate vorticity are the most generic flows, called non-homentropic. These  are the flows we consider. Such flows are baroclinic. They should be contrasted  with  homentropic flows where vorticity is degenerate and \(\Omega\wedge\Omega=0\). Barotropic flows, where the gradient of pressure and particle number density are collinear, are homentropic. They occur if the energy density in \eq{66} does not depend on the spectator density.  This is why we must be concerned about the neutralized spectator medium.  Would the energy density in \eqref{66} be  a function of the particle number only,  the equation of state \eqref{171} would prevent the components of the 4-momentum from being independent, prompting the vorticity 2-form to be degenerate.  
% It endows the 4-dimensional momentum space with a symplectic structure. 

When the vorticity 2-form is non-degenerate, the fluid phase is the cotangent bundle of the group of spacetime diffeomorphisms times the gauge group. It is spanned by \underline{five} five independent field:  four independent  components of the 4-momentum \(\pi_\mu\)  and, the chiral phase \(\Theta\). The five independent fields reflect  five independent conservation laws:   the energy, the spatial momentum, and for the particle number \cite{W}.    %We already mentioned that from a semiclassical perspective, the phase space is compact. Specifically,
% in the  chosen units the 
% {\it theta-angle} multiply covers a circle  \(0\leq\Theta<2\pi\). A change
%of \(\Theta\)  by \(2\pi\) along a time-like direction
%alters the action \eq{101} by \(2\pi N\) and  keeps \(e^{i{\cal S}}\)
%single-valued since the number of particles   \(N\) is an integer.   Compactness is
% consistent with  Onsager's semi-classical quantization of circulation \cite{Onsager}: 
%\(\oint \pi_\mu dx^\mu=2\pi\times\text{integer}\)
%in units of \(\hbar\). This suggests to use the Planck constant
%\(\hbar\) as a unit of momentum. 
\\
\paragraph{5. Theta terms.}
After these initial steps, we proceed to the modification of the Hamilton
functional \eq{11}. We  add a combination of two  theta-terms. One is based on the Pontrygin form defined by \eq{20}, another is the spacetime first Pontryagin form 
\begin{multline}
\Lambda=\Lambda_0+\tfrac k 2\Lambda_\Theta \,,\\
\Lambda_\Theta[\pi]=\tfrac {1}2  \tfrac 1{(2\pi)^2}\int\Theta\,[\Omega\wedge
\Omega+\tfrac
{1} {24}\, \tr(
 R\wedge R)]
\,.\la{271}
\end{multline}
Here \(k\) is a parameter.   These terms are  distinct.
 The former is constructed from the dynamical field of the  fluid  momentum and was previously appeared in Ref. \cite{W}, whereas the latter depends solely on the geometry of spacetime and is not dynamical \cite{FootnoteR}. 
This form is chosen based on the following arguments: (i) we do not want to extend the fluid phase space; (ii) we do not want to change the equation of state \eq{171}. In order for this to happen, the added term must be topological in the sense that it does not depend on the metric. In other words, it should be written solely in terms of differential forms. These are strong requirements which allow only for a choice in coefficients, but even this is restricted as discussed in par. 8. 
  \\
\paragraph {\it 6. Cocycle, the index of the Dirac  operator and relation to fermions.} 
The significance of the added term in \eq{271} becomes clearer through its connection with the spectral determinant and the index of the Dirac operator.   

We consider the Dirac operator 
\(\slashed{D} = \gamma^\mu D_\mu\), acting on a state of definite momentum \(\pi_\mu\), where \begin{equation}
D_\mu = \partial_\mu - i \pi_\mu - \tfrac{1}{2} \omega_{\mu\alpha\beta} \gamma^{\alpha\beta}\,.
\end{equation}
Here, \(\omega_{\mu\alpha\beta}\) denotes the affine spin connection, and 
\(\gamma^{\alpha\beta} = \tfrac{i}{4} [\gamma^\alpha, \gamma^\beta]\) is the spin operator.

We also consider the Weyl (right-chiral) component of the Dirac operator, defined as
\begin{equation}
\slashed{D}_R = \tfrac{1-\gamma^5}{2} \, \slashed{D} \, \tfrac{1+\gamma^5}{2}\,.\label{24}
\end{equation} 
 The Dirac operator possesses zero modes. Removing  them, we define a spectral determinant \(\det{'}\slashed{D}_R[\pi] \) as a regularized product of non-zero eigenvalues.  The determinant of the Weyl operator  \eq{24}is known not to be invariant under a  transformation  \(\pi_\mu\to\pi^\Theta_\mu:=\pi_\mu+\p_\mu \Theta\). Rather, it transforms as 
 \begin{align}
\det{'}\slashed{D}_R[\pi^\Theta]=e^{ i\Lambda_\Theta}\det{'}{\slashed{D}_R[\pi]}\,.\la{242}
\end{align}   
The  functional \(\Lambda_\Theta\) is referred to as a 1-cocycle \cite{Faddeev}. It is a multivalued functional in a sense that
the global  transformation \(\Theta\to \Theta+2\pi\) changes it by   half of the {\it index}  of  the Dirac  operator   \begin{align}
\Lambda_{\Theta+2\pi}-\Lambda_\Theta=-\pi\,\text{index}\,(\slashed{D})\,.\la{30}
\end{align} 
[We recall that the index of the Dirac operator  is the difference between the number of normalized
zero modes of \(\slashed{D}\) of  opposite chirality.]  

The index is  a  topological invariant generally taking  integer values, but in the case relevant to our setting, taking even integer values.  It equals
the sum of the  momentum Pontryagin number and spacetime Pontryagin number  \cite{Eguchi_Diff} 
\begin{align}
\text{index}\,(\slashed{D})=-  \tfrac 1{(2\pi)^2}\int [\Omega\wedge\Omega+\tfrac 1{24}
\tr\left(
 R\wedge R)\right]\,.\la{28}
\end{align}
 The explicit form of the cocycle  is identical to the functional  \eqref{271}. 
 This relation indicates that the fluid is comprised of \(k\) species  of chiral fermions. The spacetime Pontryagin form in (\ref{28}) represents  the effect of spin.
  
 As we discussed in the previous paragraph, the room  for maneuver with the  functional \eq{271} is limited. The only conceivable generalization is an integer factor  in front of the spacetime Pontryagin form, which  corresponds to ? value of spin other than 1/2.

[As a side comment, we briefly discuss a fermion representation of the fluid. We model  chiral fermions as Dirac spinor  composed of the right-handed Weyl spinor  \((\psi,0)\) and  the left-handed Weyl spinor  \((0,\chi)\), and assume that the latter  has neither inertia nor spin. Then \(\chi\) represents  the spectator fermionic medium. If  there are  \(  k\) identical species of
fermions, then the Lagrangian  density of such a model is 
 \begin{align} L=i\sum_{j=1}^k\bar \psi_j \slashed{D}_R\psi_j+i\sum_{j=1}^k\bar \chi_j  \slashed{D}_L\chi_j\,,\nn
 \end{align}
 with  \(   \slashed{D}_L:= \tfrac{1+\gamma^5}{2}(\gamma^0\p_0+i\slashed{ A}) \tfrac{1-\gamma^5}{2}\) and \(  \slashed{D}_R \) is given by \eq{24}.
 
 When we add pressure \(P(\sqrt{-p_\mu p^\mu})\), understood as a given function of the momentum, the model will describe interacting fermions flowing over  reservoir of spectator particles with the opposite chirality. Then, integrating over fermions we obtain our  fluid model with the particle currents \(n^\mu=\sum_j \bar \psi_j\gamma^\mu\psi_j\), and the spectator current \( \bar n^\mu=-\sum_j\bar\chi_j \gamma^\mu\chi_j\). ]
\\
\paragraph{7. Chern-Simons form and fluid helicity}
It is helpful to express the  functional  \(\Lambda_\Theta\)  \eq{271}  in terms of the fluid 
helicity defined as  the momentum 3-form \begin{align}
 \sigma^\mu:=\tfrac{1}{(2\pi)^2}\e^{\mu\nu\lambda\sigma}p_\nu\p_\lambda
p_\sigma\,.\la{spin}
\end{align}
In the external gauge field helicity includes 
 the  effect of Larmor precession effect \cite{FootnoteK,abanov2022axial,W}:
\begin{align}\Sigma^\mu=\sigma^\mu+\tfrac{2}{(2\pi)^2}
%p_\nu(\p_\lambda p_\sigma+
p_\nu{}^\star F^{\mu\nu}
%)
\,.\la{91}
\end{align} 
Here  \(  ^\star\* F^{\mu\nu}=\tfrac 12 \epsilon^{\mu\nu\lambda\sigma}F_{\lambda\sigma}\)  is the dual field tensor. 
 Later, in par. 10 we argue that the fluid helicity given by  \eq{spin} appears to be a fluid spin.
In terms of differential forms, \(\Sigma^\mu \) is the Hodge dual of  the helicity  3-form 
\begin{align}\Sigma=\tfrac{1}{(2\pi)^2}p\wedge( dp
+2F)\,.\la{92}
\end{align} 
With the help of the identity
%\begin{align}
\(d\Sigma=\tfrac 1{(2\pi)^2}[\Omega\wedge\Omega-F\wedge F]
%\end{align}
\), 
we obtain \begin{multline}
\Lambda_\Theta=  -\tfrac 12 \int\Theta\,[d \Sigma +\tfrac 1{(2\pi)^2}(F\wedge
 F+\tfrac
{1} {24}\, \tr(
 R\wedge R)]=\\
 \tfrac 12 \int  (\Sigma\,+K
+\tfrac {1}{24} K_G)\,d\Theta\,,\la{26}
\end{multline}
where \(K=\tfrac 1{(2\pi)^2}A\wedge dA\)
is the \(U(1)\) Chern-Simons form and \(K_G\) is the gravitational Chern-Simons form given by \eq{109} (and defined up to the exact form).

Now we turn to  the equations of motion. We begin with  
the continuity equation.

\paragraph{8. Continuity equation and the chiral anomaly.}    
A transformation of  the Hamilton functional  \(d\Theta\to d\Theta+d\delta \Theta\) at a fixed \(\pi\) gives the conserved Noether current.     We denote it by  \(I^\mu\). The contribution  of \(\Lambda_0\)  gives \( n^\mu\), as it follows from (\ref{171}). The contribution of \( \Lambda_\Theta\) is conveniently read from \eq{26}. Summing these contributions we obtain
\begin{align}
I^\mu=j^\mu+\tfrac {k}2  \left(K^\mu +\tfrac 1{24} K_G^\mu\right)\,, \la{4}
\end{align}
where \(  j^\mu\) (commonly referred  as a vector current) is 
\begin{align} j^\mu=n^\mu+\tfrac
k2 \Sigma^\mu\,.\la{81}\end{align}
The first term in this formula is an electric charge carried by the flow, the second term is neutral. It was argued in \cite{W} that it represents the spin \(\sigma_{\alpha\beta}\) of the flow (see, par. 9 for further discussion).

Eq.\eq{4}
illustrates the meaning of the chiral  anomaly:   the conserved Noether
current  is  neither gauge-invariant nor diffeomorphism-invariant; however, its divergence is.
The non-invariant parts are the Chern-Simons forms:  \(K^\mu\)  and  $K_G^\mu$.   They signify that our system 
is coupled with a reservoir capable
 of supplying and swapping particles and flipping their spin by changing the fluid chirality. One may interpret  the gauge invariant vector field \(j^\mu\)  as a  part  of the current that flows through the fluid, while
 the non-invariant part  \(\tfrac {k}2  \left(K^\mu +\tfrac 1{24} K_G^\mu\right)\,\)
is the part of the  current that runs through a reservoir.
Their sum \(I^\mu\) is conserved  \begin{align}
\p_\mu I^\mu=0\,.\la{3}
\end{align} 
This yields the formula for the divergence of
\(j^\mu\),  commonly referred to as the chiral anomaly  \cite{Current_algebra,tHooft,Kimura,Delbourgo-Salam,Eguchi-Freund}:
\begin{align}
\p_\mu j^\mu=-\tfrac {k}2 \tfrac 1{(2\pi)^2}\,\left[F_{\mu\nu}{} ^\star\* F^{\mu\nu}
+\tfrac {1}{24} \tr\left(
{R}_{\mu\nu}{} ^\star\! R^{\mu\nu}
\right)\right]\nn\,.
\end{align}
where  we use the dual tensor  \(  ^\star \!R^{\mu\nu}=\tfrac 12 \epsilon^{\mu\nu\lambda\sigma}R_{\lambda\sigma}\).

This equation, being written in Eulerian terms,  yields a modification of the usual  continuity equation \eq{c} \begin{align}
\p_\mu n^\mu\!=-\tfrac {k}2\tfrac 1{2(2\pi)^2} [\Omega_{\mu\nu} {}^\star \Omega^{\mu\nu}\!+\!\tfrac
{1}{24} \tr\,(R_{\mu\nu}
{}^\star\! R^{\mu\nu})]\,.\la{31}
\end{align}

When integrating Eq. (\ref{31})  over a time-like spacetime cylinder
\(M^3\times [-\infty,\infty)\),  the left-hand side  of \eq{31} gives 
the number of particles swept by  the reservoir   $\Delta N=\left(\int_{M^3} n\right)|^\infty_{-\infty}$
over  the time interval \([-\infty,\infty).\)  The integral of the right-hand
side  of \eq{31} is  the
index of the Dirac operator \eq{28}
\begin{align}
\Delta N=\tfrac k {2}\text{index}\,(\slashed{D})\,.\la{40}
\end{align}
Since \( \Delta N \) is an integer and the index is an even number, the parameter \( k \) is necessarily an integer.

Two processes contribute to particle production \(\Delta N\): vortical instantons \cite{W} and gravitational instantons \cite{Eguchi_Hanson}. These are configurations which give nonzero values to the index. Unlike gravitational instantons, `vortical instantons' have a clear interpretation \cite{abanov2022axial}: The momentum  Pontryagin number given by the first term in \eq{28} is twice the change in the linking number of vortex loops. Consequently, a change of the linking by one causes an exchange of \(k\) particles with the reservoir.

We remark that the topological invariants, the momentum Pontryagin number \(P_1(\Omega)=-\tfrac 12\tfrac{1}{(2\pi)^2}\int \Omega\wedge\Omega\), determine the particle production \eq{40} via \eq{28}. They characterize classes of flows in a manner analogous to the theta-vacua states of quantum field theory \cite{ThetavacuumCallan,ThetavacuumJackiw}, with the chiral phase \(\Theta\) being analogous to the {\it axion} of the CP-problem \cite{axionWilczek,axionWeinberg}.
\\
\paragraph{9. The Euleri equation.}
Now let us discuss   the effect  of the theta-terms on energy-momentum  conservation laws \eq{241}. For that, we have to compute the action of diffeomorphisms on  \( \Lambda_\Theta \) at a fixed \( \Theta \). 
The transformation of \(\Lambda_\Theta\)  at a fixed \(\Theta\) is opposite
to  the
transformation of \(\Theta\) at a fixed \(\Omega\) and \(R\). Since \( \Theta \) is a scalar,  the transformation \(\delta_\epsilon\Theta=
\epsilon^\mu\p_\mu\Theta\) generates the  Noether current \( I^\mu\).    Therefore,  the net result amounts to  the replacement of \(\p_\mu n^\mu\) in \eq{211}  with \(\p_\mu I^\mu\), which nulls on-shell.  We end up  with the conservation laws \eq{241} with the momentum-stress-energy  tensor which retains the same form as that for a perfect fluid \eq{16}. We repeat it here
\begin{align}
&\p_\mu \tensor{T}{^\mu_\nu}=-n^\mu F_{\mu\nu},\quad \tensor{T}{^\mu_\nu}=n^\mu p_\nu+P\,\tensor{\delta}{^\mu_\nu}\,.\la{2412}
\end{align}
Together with the modified continuity equation \eq{31} they provide the complete set of equations of motions.

This result is expected. The added term 
 \eq{271}  is of a topological
 nature and should not affect the form of the stress tensor. 
However, the Newtonian form of the Euler equation is  affected.
Computing the divergence of the stress tensor \eq{2412}, we move the `rocket term'  \( p_\nu(\partial_\mu n^\mu) \) to the right-hand side:
\begin{align}
n^\mu\partial_\mu p_\nu+\partial_\nu P=- p_\nu(\partial_\mu n^\mu)\,\label{E}
\end{align}
and treat it as a reactive force caused by particle production. It includes the divergence of the  particle number current \(  \partial_\mu n^\mu\), which is no longer zero and is given by the modified continuity equation \eq{31}. 

The advection variables \(  \Theta\) and \(  \bar n\) disappear from the conservation laws (\ref{31},\ref{2412}) as expected.
\\
\paragraph{10. Spin and the chiral curvature effect.}

Because our fluid is composed of Weyl fermions, it possesses spin, and
the spin interacts with  the fluid vorticity and the
spacetime curvature. To highlight this effect, we omit
the external gauge field. 

Using the continuity equation in the form of \eq{31}, along with the identity
\[
4\Omega_{\mu\nu}{}^\star \Omega^{\lambda\mu} = \delta^\lambda_\nu \Omega_{\gamma\mu}{}^\star \Omega^{\mu\gamma}
\]
(which holds for any antisymmetric 2-tensor), and the first Bianchi identity for the Riemann tensor, we express the Euler equation  \eq{E} as:
\begin{align}
n^\mu\partial_\mu p_\nu+\partial_\nu P = k \left( \sigma^\mu \Omega_{\mu\nu}
+ \tfrac 12 \sigma_{\mu\alpha\beta} \tensor{R}{^\alpha^\beta^\mu_\nu}
\right)
\,,\label{35}
\end{align}
where
\begin{align}
\sigma_{\mu\alpha\beta} = \tfrac 1{(2\pi)^2}\left(p_{[\mu}\partial_\alpha p_{\beta]}
+ \tfrac 1{2 \cdot 24} \, p_\lambda
{}^\star\tensor{R}{_\alpha_\beta_\mu^\lambda}\right)\,,\label{33}
\end{align}
and \( \sigma^\mu \) is the fluid helicity given by \eq{spin}, with \( [\dots] \) denoting antisymmetrization.

The last term on the RHS of \eq{35} is the fluid analogue of the Mathisson-Papapetrou force acting on a spinning particle moving in curved spacetime \cite{Papapetrou}. Pursuing this analogy, the totally antisymmetrized tensor \( \sigma_{[\mu\alpha\beta]} \) is identified with the spin current (by virtue of the first Bianchi identity, only the antisymmetric part contributes to \eq{35}). The spin current given by \eq{33} is totally antisymmetric. This is consistent with fermions, as the fermionic spin current \( \sigma_{\mu\alpha\beta}=\bar{\psi} \gamma_\mu \gamma_{\alpha\beta} \psi \) is also totally antisymmetric. Following the terminology of Ref.\cite{Kharzeev}, one may refer to the curvature-dependent part of the spin current in \eq{33} as the \textit{chiral curvature effect}, a geometric counterpart of the chiral magnetic effect \cite{FootnoteK}.

The spin current defines the spin density matrix as the projection of the spin current onto the flow direction:
\[
\sigma_{\alpha\beta} := -u^\mu \sigma_{[\mu\alpha\beta]}.
\]
Observe that the curvature-dependent contribution vanishes from the spin density, as it is solely expressed in canonical terms of the flow, independent of the background. Then, the fluid helicity given by \eq{spin} and entering \eq{91} and \eq{35} is the dual of the spin density:
\[
\sigma^\mu = \tfrac 12 \varepsilon^{\mu\nu\alpha\beta} u_\nu \sigma_{\alpha\beta}.
\]

This leads us to a noteworthy result: the spin of our fluid is identical to the fluid helicity, \textit{acquired through motion}. It is not an inherent or constitutive property of the fluid. For this reason, spin does not appear in the momentum-stress-energy tensor \eq{2412}. Consequently, our equations of motion \eq{35} differ from those of the conventional spinning fluid (see, e.g., \cite{Obukhov}).

Alongside the spin-curvature interaction (the Mathisson-Papapetrou type force), the first term on the RHS of \eq{35} represents a force arising from a spin-vorticity interaction. Both forces are of geometric origin, stemming from the anomaly contribution to the continuity equation \eq{31}.

For reference, we quote a tensor  representation of   the Mathisson-Papapetrou  force
\[
\,\nabla_\mu\tau^{\mu\nu} = \xi\, p^\nu \tr\left({R}_{\mu\lambda}{} ^\star\! R^{\mu\lambda}\right)\,,
\]
where we abbreviate \( \xi := \tfrac 1 2 \tfrac 1{(2\pi)^2} \tfrac 1{24} \). The tensor is asymmetric \( \tau = \tau_S + \tau_A \) with  symmetric and antisymmetric components  given by:
\begin{align*}
& \tau_S^{\mu\nu} = -\tfrac \xi 2 \nabla_\alpha \left( p_\beta \left( {}^\star \tensor{R}{^\alpha^\mu^\beta^\nu}
+ {}^\star \tensor{R}{^\alpha^\nu^\beta^\mu} \right) \right)\,,\\
& \tau_A^{\mu\nu} = \xi\nabla_\alpha \left( p_\beta {}^\star \tensor{R}{^\mu^\nu^\beta^\alpha} \right)\,.
\end{align*}
The symmetric part appeared in Ref.\cite{JPi}; the antisymmetric part is the divergence of the curvature-related component of the spin current.
\\

Summing up, the Chern-Simons modification of the perfect fluid  presented here is consistent with the kinematics  of Weyl fermions and incorporates both the chiral current and the chiral  gauge-gravitational anomalies characterized  the chiral fermions. This fluid is necessarily spinning, with its spin equal to the fluid helicity modified by the chiral curvature effect, and it exerts a combined force representing a geometric spin-orbit and spin-curvature interaction.
\\

\begin{acknowledgments}The work was supported by the NSF under Grant NSF
DMR-1949963. The author
gratefully acknowledges A. G. Abanov and A. Cappelli for their collaboration
on this subject and G. Volovik for his useful comments.
  \end{acknowledgments}

%{\it Note added}: After completing this work, we became aware of an earlier, noteworthy  paper by Nair, V. P., Ray, R., and Roy, S., {\it Fluids, anomalies, and %the chiral magnetic effect: A group-theoretic formulation}, Phys. Rev. D {\bf 86}, 025012 (2012) \cite{Nair}. This paper addresses topics closely related to %those discussed here and likewise employs the canonical approach to fluid mechanics used in the present work. We thank the anonymous referee for making %us aware of this work.


\begin{thebibliography}{10}
\bibitem{FootnoteR}  The fluid Hamilton functional modified by \(\Theta R\wedge R\) and its implication to the  continuity  equation given by the last term in  \eq{31} was introduced in Ref. \cite{Nair} and \cite{Abanov_Cappelli}.
\bibitem{FootnoteE}  The   density of the reservoir per particle 
 \(\overline n/n\), a  scalar, is analogous to the specific entropy in  traditional fluid mechanics.  For a view on entropy as a
 as a spectator medium, see  \cite{Carter}.
\end{thebibliography}
\end{document}